# Drivers and Challenges of Internet of Things Diffusion in Smart Stores: A Field Exploration

Michael Roe[a], Konstantina Spanaki[b*], Athina Ioannou[c], Efpraxia D. Zamani[d], Mihalis Giannakis[b]

[a] School of Business and Economics, Loughborough University, Loughborough Leicestershire United Kingdom

[b] Audencia Business School, Nantes, France

[c] Surrey Business School, University of Surrey, Guildford, United Kingdom

[d] Information School, the University of Sheffield, Sheffield, United Kingdom

[*]*Corresponding email address:* [kspanaki@audencia.com](kspanaki@audencia.com)

# Drivers and Challenges of Internet of Things Diffusion in Smart Stores: A Field Exploration

## Abstract

The digitally disruptive environment has evolved rapidly due to the introduction of new advancements within the field of smart applications. Applications of one of the most prominent technologies, Internet of Things (IoT), often appear in the retail sector, where smart services have transformed the customer experience holistically. Presented in this paper are the findings from an exploratory field study in the retail service sector, which drew on the views of experienced practitioners about the smart store experience and the associated changes. The paper presents an overview of the drivers of smart retail service diffusion and the relevant challenges, such as the business expectations and the heterogeneity of devices. The arising themes indicate that IoT security is a major challenge for businesses installing IoT devices in their journey towards smart store transformation. The paper highlights the importance of a secure data-sharing IoT environment that respects customer privacy as the smart experience in-store offers data-driven insights and services. Implications for research and practice are discussed in terms of the customer experience relevant to the identified challenges.

## 1 Introduction

The cost-effective, accessible nature of Internet of Things (IoT) devices, combined with their ability to connect a business firm to both its environment and its customers in real time, has made the technology highly attractive to a wide variety of industry sectors (Metallo et al., 2018). It has in fact been identified as one of the four leading disruptive technologies that will revolutionise the retail industry (Grewal et al., 2017).

At the same time, customer expectations have begun shifting from being product-centric to being more experiential (von Briel, 2018). The exponential development of IoT makes it essential for catering to the quality expectations of end users and for monitoring the processes in a business firm. However, focusing on the experiential aspects requires the collection of unprecedented amounts of data and the use of advanced analytics (Bradlow et al., 2017). Given the customer-centric dynamic of a service environment, the potential volume of personal data that can be amassed is vast, thus having obvious implications for data privacy (Aloysius et al., 2018; Inman & Nikolova, 2017a). Broader challenges, including security concerns, also need to be considered (Marikyan et al., 2020). From a practical perspective, it is widely acknowledged that the simple increase of devices within a network poses a threat because it increases the exposure to potential attacks (Jing et al., 2014; Roman et al., 2011). The heterogeneous nature of IoT devices further increases the risks because it raises the degree of complexity of the security requirements as IoT introduces computationally weak devices in an online environment, which contributes to system vulnerability (Jing et al., 2014; Roman et al., 2011).

The recent advances in sensor networks and IoT and their widespread adoption and diffusion have helped facilitate monitoring and quality control. However, the translation of traditional security protocols onto an IoT system is inappropriate due to the differences between an IoT infrastructure and a 'traditional' computer network. Be that as it may, while there have been many academic studies concerning IoT (and IoT security) at both the conceptual and low technical levels, the studies addressing IoT at the system level are relatively sparse (Boyes et al., 2018; Dijkman et al., 2015).

Seeking to address this gap, the present study explored the drivers of IoT implementation in smart stores, the relevant challenges posed by such and the implications of these for the customer service experience. We addressed our research question *(What are the drivers and challenges of IoT implementation in smart stores, and which of them are relevant to customer-facing services?)* by drawing from the existing frameworks of service systems and from the diffusion of innovation (DOI) theory. With regard to method, we used the field study design and combined a literature review with interviews of practitioners from the retail sector with expertise in implementing technology projects. Our findings contribute to the IoT literature by identifying the security facets that are specifically relevant to IoT in the retail service industry. We also offer practical implications that take the form of an agenda providing feasible potential solutions for IoT-related challenges, emphasising the need to tailor these to meet the requirements of a customer-facing service environment.

## 2   Prior Research

Over the past 20 years, the aim of IoT diffusion has gone beyond simply making IoT a technology that pervades multiple aspects of modern life (Lee & Lee, 2015). The ubiquitous nature of IoT is discussed in the seminal paper by Atzori et al. (2010), which states that IoT enables various objects to 'interact with each other and cooperate with their neighbours to reach common goals'. The same authors explain that RFID (radio frequency identification) will be a key technology in IoT moving forward, and that sensor networks, combined with RFID technologies, will further enable the digitalisation of the real-world environment. Given that IoT leads to the generation of an exponentially larger amount of data compared to the traditional web-based technologies, utilising cloud technology is the only feasible way of storing, accessing and analysing data in a useful way in relation to IoT (Gubbi et al., 2013). While there are challenges surrounding synchronisation and standardisation between different cloud vendors, the cloud can still have the potential to manage the big data generated from IoT if the reliability of IoT cloud-based services will be ensured and if these services will be validated and managed well (Al-Fuqaha et al., 2015).

Overall, the advancement of IoT can enhance the quality of the everyday customer experiences (Whitmore et al., 2015). However, as this technology becomes more sophisticated and pervasive, the customers' private and sensitive data are collected and shared extensively with known or unknown entities, often without the customers' knowledge (Wei et al., 2014; Yun et al., 2019). This challenge can also be seen in the retail industry and services, where private data are collected through IoT devices to provide customers with a tailored experience.

In the following sections, we explore previous studies on the IoT applications in the service industry and smart stores and the challenges of IoT diffusion. Then we examine the theoretical background of IoT technology adoption and diffusion for the service industries and smart stores to frame the research agenda for our field study.

## 2.1 Internet of Things and Customer Journey in Smart Stores

The World Economic Forum identified IoT as one of the eight technologies expected to disrupt the retail industry in the near future (Accenture, 2017). In the potential store of the future, an IoT-enabled environment will facilitate moving from product centricity to an 'experiential' customer journey (Grewal et al., 2017). From a business perspective, IoT can also facilitate radical advantages through real-time instrumentation as the automation and optimisation of manual tasks can reap massive efficiency benefits especially within industrial processes (Gierej, 2017). The simple reduction of hardware costs for a business firm's infrastructure and the generation and use of big data are considered the main drivers of investment in IoT. From the perspective of individual customers, IoT creates value for them through its ability to predict and address their needs in real time. IoT's ability to make products remain current and up to date, and to generate meaningful data that can be used to enable personalised services, support a 'path to profit' focused on IoT's ability to stimulate recurring revenues by fostering closer relationships between businesses and their customers (Metallo et al., 2018).

The service industry has been radically transformed due to the emergence of IoT applications. More specifically, in the retail industry, IoT is streamlining and automating processes that revolutionise services and the overall customer/shopping experience, introducing significant and simultaneous benefits for both consumers and businesses (Giebelhausen et al., 2014a). IoT-based technologies can provide personalised promotions to customers to manipulate their path through the store (Hui et al., 2013) and to induce a rise in the value of a customer's basket. Another facet of personalised shopping demonstrates that encouraging shopping on a mobile phone reduces the need for blanket discounts, which overall reduce the company's costs (Wang et al., 2015). In other studies, the use of big data analytics to control in-store pricing showed that for a 100-store enterprise, the increase in profits as opposed to human pricing control could be up to US$11 million (Bradlow et al., 2017).

For business firms adopting IoT, superior customer experience and supply chain optimisation and innovations in in-store experiences can be achieved through the technology, resulting in higher efficiency and profitability for the business (Gregory, 2014). Data from IoT sensors, such as environmental and motion data, enable retailers to offer personalised, tailored customer experiences by monitoring the store traffic and customer demand in real time, allocating assistants where they are most needed or adjusting the store layout, increasing the store management efficiency (in smart stores where the stock of products is being updated in real time) and monitoring and predicting the in-store waiting times. Although not necessarily linked to IoT, sensor networks support the collection of big data, which help better characterise an environment (whether physical or social). However, sensor-enabled solutions such as smart shelves and robotic assistants can monitor a store's performance in real time and are among the most prominent examples of IoT applications in the retail industry showcasing the unique nature of IoT applications in the service industry (Intel, 2017).

Technological changes and implementations in the retail industry, such as IoT, have significantly influenced consumer decision making (Hamilton et al., 2021). For consumers, the rapid diffusion of IoT in the retail sector has radically transformed customer experience, and more specifically, the customer journey (Hoyer et al., 2020) in all of its phases, from the pre-purchase phase to the purchase and post-transaction phases (Lemon & Verhoef, 2016). In the pre-purchase phase, aspects such as smart trolleys, smart mirrors and interactive fitting rooms can transform the customer experience, offering a more immersive and personalised approach (Ogunjimi et al., 2021; Shankar et al., 2021). New and innovative touchpoints have been introduced while older ones have been redeveloped to further enrich customer experience and create new value (Hoyer et al., 2020). Customers can be identified the moment they enter the store through beacon technology and can receive personalised notifications and recommendations through their smart devices on the basis of their purchase history and personal preferences. Aiming to revolutionise the transaction stage of the customer journey, retailers have been incorporating several in-store disruptive IoT touchpoints that can enhance customer convenience and increase customer satisfaction, such as by decreasing customer queuing or eliminating it altogether. From scanning the products on their own (e.g., Zara's self-check-out) and paying via smartphone or wearable device, the purchase phase of the customer journey now includes walking out of the store with no checkout process at all, such as via Amazon Go (Shankar et al., 2021).

Overall, it becomes apparent that the implementation of IoT in smart stores is radically transforming the customer experience, with various new touchpoints being created and others being reconfigured (Hoyer et al., 2020). Such transformation may exert a significant influence on other related aspects and follow-on consumer experiences, interplaying in the customer journey, from customer satisfaction with the company performance and service quality to trust in a company and customer engagement (Lemon & Verhoef, 2016). However, while there is existing evidence that IoT services in the retail sector have a positive impact on customer satisfaction and experience (Ratna, 2020), the research in this area is still in its infancy. Considering that the customer experience is a multi-dimensional concept relating to '… cognitive, emotional, behavioral, sensorial, and social responses to a firm's offerings…' (Lemon & Verhoef, 2016, p. 74), it is imperative to examine and realise the profound impact of IoT throughout the whole spectrum of the customer journey and the different dimensions of customer experience.

## 2.2 Challenges of Internet of Things Diffusion in the Retail Sector

IoT presents numerous opportunities in the organisational environment and has the potential to revolutionise the way multiple industries operate. However, challenges in securing, verifying and storing data also exist, and these challenges act as a barrier to the more widespread adoption and diffusion of IoT. Some of these challenges are prevalent for IoT in general within a customer-facing industry. Nevertheless, the added dimension of the personal nature of the collected data has further security ramifications.

Firstly, with regard to the vast amounts of data generated and their management, an issue that arises pertains to how data are stored. Then, other questions pertaining to how 'quality' data are identified, isolated and prioritised also arise (Lazer et al., 2014a). Secondly, there is the issue of the mixed-media format of the data being transmitted and analysed, and the infrastructural complications stemming from the processes employed for transmission and analysis. Within the hyperconnected and hyper-accelerated innovation cycle that exists within the technology sphere, there is a potential for the advancements to become chaotic, especially without concrete and universal regulations in place (Weber & Studer, 2016a). The issue of security can also be raised at the device level, which is intrinsically linked to the most critical issue of data privacy (Palattella et al., 2016a).

Several key factors create a bespoke challenge for IoT security and privacy: device heterogeneity, data heterogeneity and low-power device nature (Al-Fuqaha et al., 2015; Atzori et al., 2010; Gubbi et al., 2013). These factors, combined with the increased number of devices within an IoT network, point to the supreme importance of security for the successful dissemination of IoT.

Specifically, in the data-driven IoT environment of smart stores, the customer journey is based on the insight and personal information of each customer for a tailored shopping experience

(Hu et al., 2018). Services are based on the information shared by the customers through the IoT platforms, and in a fully smart environment, customers have to provide their consent to divulge certain personal information to companies and deny access to other information (Brous et al., 2020). In the grand vision of IoT, where all devices are interconnected, data control systems must be able to accurately control what data can be transmitted and to whom (Brous et al., 2020).

IoT faces specific challenges with regard to privacy. IoT use is not yet sufficiently regulated to ensure privacy and security for customers (Hu et al., 2018). Due to the heterogeneous nature of IoT, this issue needs to be addressed from different perspectives (Hu et al., 2018; Lu et al., 2018).

A study about consumer-facing retail technologies found that customers are generally highly supportive of technologies that reduce queuing times but are uncomfortable with proximity marketing (Inman & Nikolova, 2017). This indicates that accepting 'help' from technologies for the adoption of 'privacy-invasive' technologies constitutes a cultural shift. This can be aided, as discussed earlier, by companies making concerted efforts to show that customer privacy is their priority.

While the literature reviewed in this section pertaining to IoT privacy and security at the generic concept level comes from an abundant research bank (Al-Fuqaha et al., 2015), there is still more ground to be covered in the field of IoT adoption and diffusion with regard to data-sharing and security challenges (Brous et al., 2020; Hwang et al., 2015; Pauget & Dammak, 2019). Our study focused on this gap and highlighted the importance of IoT security and privacy in the customer-facing environment of smart stores.

## 2.3 Theoretical Background

The present study sought to address the gap in the existing knowledge regarding the drivers and challenges of IoT implementation, and to offer potential solutions particularly for customer-facing service environments and smart stores within the retail sector. Therefore, the theoretical framing of this study drew from three streams of theory: the DOI theory (to provide details regarding the diffusion of IoT as a technological innovation), IoT network background (to account for the IoT-specific aspects) and the service-dominant (SD) logic (to explore the specifics of service systems and the diffusion of IoT in smart stores or retail services).

There have been few studies on the secure implementation of an IoT system in a real-world environment (Metallo et al., 2018) considering the security features within a retail environment. Seminal studies have provided instructions on what to consider when setting up an IoT system, but they have not covered all the challenges that may be posed by IoT implementation (Goad et al., 2020). The framework for industrial IoT (Boyes et al., 2018) can be used in the planning phase, with a focus on security, to ensure that there will be no intrinsic system flaws. While the framework provides valuable questions for considering the attack surface of an IoT system, practical solutions are not supplied.

To understand how new technologies are disseminated throughout society, the DOI theory (Rogers, 2003) presents a useful conceptual framework. The DOI framework presents the requirements for understanding how a new application technology (in this case, IoT in a customer-facing environment) can be evaluated in terms of its successful proliferation and the subsequent challenges that may be faced during its implementation. The DOI theory presents five variables that determine the adoption rate of innovations: perceived attributes of the innovation, type of innovation decision, communication channels, nature of the existing social system and extent of the change agent's promotion efforts. While all these five variables provide a rounded view of how likely an innovation is to flourish within an industry, the perceived attributes of innovations are the most widely considered in the literature (Rogers, 2003) and were adopted in this study.

Historically, the service industry has focused on exchanging goods, resulting in a landscape aligned with a transactional infrastructure, facilitating the exchange of tangible resources (Lusch & Vargo, 2006). However, changes have recently been seen in this landscape; it now includes intangible assets, value co-creation and relationships. This has resulted in the development of the SD logic (Vargo & Lusch, 2008), which is encapsulated in the shift occurring within the service industry and is especially prevalent in retail- and customer-centric studies (Grewal et al., 2017). On the basis of the SD logic, there is a perspective on service innovation (Lusch & Nambisan, 2015) that segments the concept into three key themes: (1) service ecosystem, the network of actors governing the landscape of the service exchange; (2) service platform, the combination of resources (both physical and intangible) forming a provision and (3) value co-creation, the actions motivating the resource integration and actor interactions within the service ecosystem. The SD logic departs from the other logics in service science by heightening the value creation process, broadening the scope of resources and supporting collaboration within and between service systems.

Combining the works of Lusch and Nambisan (2015), Rogers (2003) and Boyes et al. (2018) allowed us to formulate a theoretical framework for our field study. Specifically, we drew from the work of Boyes et al. (2018) to identify the IoT network vulnerability aspects, and from the works of Lusch and Nambisan (2015) and Rogers (2003) to identify the SD themes and the perceived attributes of innovations, respectively, which inform the holistic appraisal of the cyber-business environment. The decision to adopt a deductive design for the research, with a specific amalgamation of theories serving as a theoretical background, reflects the pragmatist standpoint of this study regarding the implementation of IoT within the retail industry setting. For a genuinely holistic appraisal, it is not sufficient to merely consider the technical aspects of security design, without an appreciation of how the technology interacts with the other elements of its environment to generate value and consider the factors that will impact its implementation and reception.

# 3 Research Design

The present study contributed to the field of evidence-based management through the critical evaluation of relevant, high-quality studies and practitioner expertise and judgement (Denyer & Tranfield, 2006; Tranfield et al., 2003). Therefore, the study was multifaceted, deriving evidence from the existing academic research and harnessing the explicit and tacit knowledge of those working in the field (Bryman & Bell, 2011). The approach corresponds to the definition of an elicitation study provided by Edgar and Manz (2017), with an exploratory qualitative field research design.

Random sampling is not appropriate in an elicitation study, where the objective is to capture knowledge from experts (Marshall, 1996; Suri, 2011). Instead, a purposeful key informant sampling technique is required (Marshall, 1996; Suri, 2011). We thus endeavoured to identify industry professionals working within a retail setting and in an area with an IoT focus. We were also open to snowball sampling, whereby we asked the respondents to suggest other professionals from their own networks who could provide interesting insights on the study topic (Marshall, 1996; Suri, 2011). In total, 10 people were interviewed, each of them providing an information-rich case where experience was the critical sampling focus. All the participants shared their expertise and experience in the multiple business firms where they had worked with IoT over the previous years. A description of our respondents is given in Table 1.

Table 1. Description of respondents

| ID | Industry | Main country of operation | Global |
|----|----------|---------------------------|--------|
| 1  | IoT consultancy | United Kingdom | Yes |
| 2  | Retail | United Kingdom | Yes |
| 3  | IoT services | United Kingdom | No |
| 4  | IoT consultancy | United Kingdom | Yes |
| 5  | IoT consultancy | United Kingdom | No |
| 6  | Retail | United Kingdom | Yes |
| 7  | Hospitality/retail | United Kingdom | Yes |
| 8  | Hospitality/retail | United Kingdom | Yes |
| 9  | Hospitality/retail | United Kingdom | Yes |
| 10 | Retail | United States (with operations in the United Kingdom) | Yes |

The views of the individuals who were selected to participate in the field study represented those of different retail firms in the UK or of consulting firms for IoT implementation (10 retail organisations/consulting firms in total). 'Experience in IoT for retail customer-facing applications' was the primary participant selection criterion. Each participant was interviewed twice (for approximately 30 minutes per interview), firstly at the exploratory interview stage and secondly at the follow-up confirmatory stage (after the themes were generated).

For the purpose of data collection, we conducted semi-structured interviews. We came up with an interview protocol on the basis of our theoretical framework, which we shared with our respondents prior to the interview as a quality measure to facilitate full responses (Patton, 2002). This approach was chosen as opposed to an 'informal conversation' or 'standardised' method because it provides a better opportunity to obtain detailed answers showing the respondents' specific knowledge. Furthermore, as the study's objective was not to compare the knowledge of professionals but to consolidate it, there was arguably a call to increase the variance rather than to reduce it (Marshall, 1996; Patton, 2002). In addition, given our pragmatist approach, we took care to ensure the identification of problems, and we encouraged the respondents to provide potential solutions to the problems that they highlighted without leading them to such solutions.

The interview guide was designed to reflect the multifaceted nature of IoT applications in customer-facing environments such as smart stores. Starting with more high-level questions concerning how and why IoT is used in industry (and specifically in customer-facing service environments), the guide then examined the specific setup within each business firm and how each participant considered IoT diffusion and the implications of such for the customers. It was then placed under a pragmatic project management lens both in the context of real-world experience and in a hypothetical scenario where 'best practice' could be observed. The interview guide was used as a basis for the conversation, but as previously explained, due to the semi-structured approach employed in the research, it was not always strictly adhered to. Instead, the guide was primarily used as a springboard for further probing, and the interviews were allowed to flow in the direction of the interviewees' expertise but were guided by the themes explored by the research.

For data analysis purposes, we adopted the thematic analysis approach to identify the thematic clusters emerging from our interviews and to capture opportunities and challenges. Braun and Clarke (2006) describe thematic analysis as 'a method for identifying, analysing, and reporting patterns (themes) within [qualitative] data'. Theme development was done via a hybrid inductive/deductive approach, which facilitated finding general conclusions from the full dataset (inductive) while ensuring that the more specific research objectives would also be explored (deductive) (Boyatzis, 1998; Braun & Clarke, 2006). Specifically, we analysed the interview data to identify the major themes emerging therefrom. Through this, we identified business considerations, devices and infrastructure, project management, privacy and data and keeping humans in the loop as the major themes encountered when implementing IoT in-store. In the second phase, we examined the themes in greater detail by iteratively reading and comparing them. We also rechecked the consistency of our coding with the aim of classifying and organising our subthemes across the interviews. This resulted in a total of 16 subthemes under the five major themes regarding the security aspects of IoT system implementation at a store. Finally, we conducted reliability analysis, summarising our findings, evaluating them, identifying relevant and representative vignettes from the interview data to illustrate the emerging themes and relating our findings to the existing literature. Table 2 encapsulates our findings from our analysis.

Hennink et al. (2017) define code saturation as the stage where no additional issues are identified from the data obtained from the respondents attained through an inductive, content-driven approach. To ensure code saturation, as we were conducting the interviews, we were continuously analysing our material to establish that no more themes were emerging. We thus stopped conducting additional interviews when we reached code saturation.

**Table 2.** Data coding structure

| Theme | Subtheme | Code |
|---|---|---|
| Business considerations | Drivers | • Customer expectations<br>• Being in the market and remaining competitive<br>• Improves the 'customer journey'<br>• Cost-effective, with better insights (market research can be cheaper) |
| | Challenges | • Heterogeneous devices (standards)<br>• Security (cost, oversight of devices' production)<br>• Affordability and operational costs<br>• Security costs have to be balanced in favour of business profits. |
| | Opportunities | • Personalised service to build brand loyalty (personalised customer journey): footfall cameras, wireless tracking, feedback, how people move within a store<br>• Align with the 'fail fast' organisation mentality.<br>• Being the first to 'crack IoT (Internet of Things)' (if 'done well') can be a unique selling point for a UK store. |
| Devices and infrastructure | Security (by) design | • Device/systems provenance (manufactured by third parties)<br>• Diverse/changing regulations and data handling<br>• Patching |
| | Device management | • 'Shadow IT [information technology]'<br>• Network access control (monitoring what is on the network)<br>• Restrict access of IoT devices to networks<br>• No access to the corporate network |
| | Heterogeneity and complexity | • No standardisation<br>• Operating costs increase<br>• Identification of 'experts' in multiple operating systems/devices<br>• Response time and protocols<br>• Training |
| Project management | Implementation | • Business case for the project – Is there a return on investment (ROI)?<br>• Project design (including an ill-suited infrastructure for an IoT-capable store - retrofitting)<br>• Procurement |

| | | - In-house device testing (taking devices apart to see how they work)<br>- Trialling: No trial equals 'firefighting' not just for security but also for usability; trial in the busiest store<br>- Design of services: Involve the security team at all the stages; needs a business case and ROI<br>- Rollout and monitoring |
|---|---|---|
| | Security lifecycle | - Ensure that it is considered from day 1 (continuous monitoring)<br>- Consideration for procurement and running costs |
| | Organisational culture | - Impacts on privacy-related perceptions<br>- Continuous observation for identifying potential concerns<br>- Change management and resistance to change |
| Privacy and data | Compliance | - Access, rights and permissions<br>- With reference to the General Data Protection Regulation (GDPR)<br>- Data storage: for how long, where, how<br>- How can customers access their data (if they request it)? |
| | Understanding data | - Not all data are needed (collect only those that are needed); cherry-pick what to process and store<br>- Where do the data come from? (Is regulation the same everywhere?)<br>- Value of data (individual and combined data streams); metadata can be used to create 'real' data<br>- Infrastructure |
| | Responsibility | - Outsourcing does not free a company from the responsibility of protecting customer data.<br>- Show customers you are serious about data handling (customer requests).<br>- Being GDPR compliant will not necessarily make the customers 'comfortable' with data collection. |
| | Supply chain | - A business firms needs to be confident that its supplier has at least the same security controls as it does over the customers' data.<br>- Hardware device manufacturers may not be concerned about data protection.<br>- Personal identifiable information is the lifeblood of retailers.<br>- Retailers are more concerned with aggregate data. |
| Keeping humans in the loop | Usability versus security | - The security systems in place should not interfere with the employees' roles.<br>- Security should be balanced (should allow employee innovation).<br>- Use the same processes as much as possible. |
| | Education | - Provide education on 'security devices'.<br>- Protect against human errors (but educate humans too). |
| | Behaviour | - Employees will find loopholes in the new system to make things work 'as they used to'.<br>- Unclear recommendations are hard to pass on to the employees.<br>- Tie employee education to the employees' experiences or day-to-day activities (make it relatable). |

# 4   Findings

In this section, we present our findings organised according to the five main themes that emerged from the empirical data we had obtained: business considerations, devices and infrastructure, project management, privacy and data and keeping humans in the loop. A summary of our findings is provided in Figure 1.

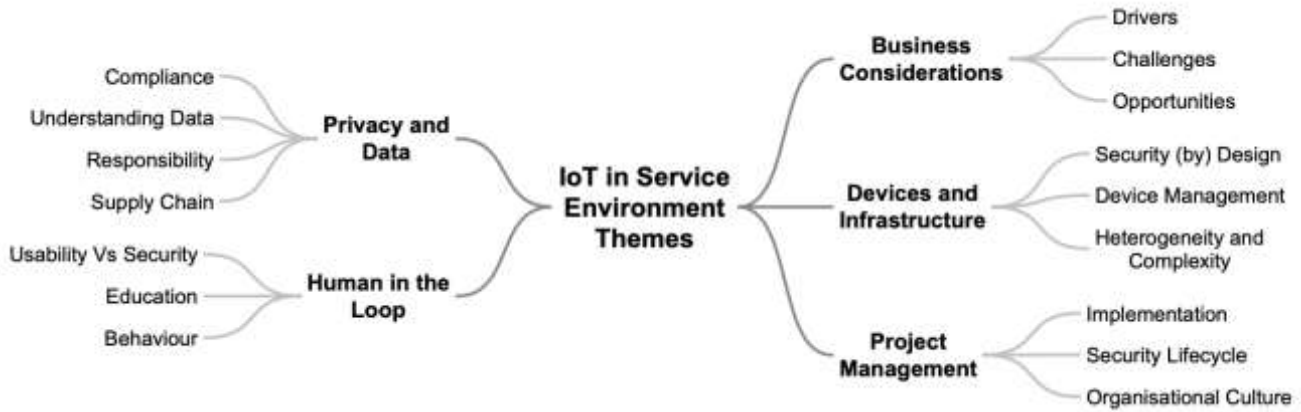

**Figure 1.** Themes and subthemes arising from the interview data

## *4.1 Business Considerations*

Business considerations emerged as a core theme, particularly 'the greatest business drivers', 'the key challenges' and the opportunities for implementing an IoT system within a customer-facing service environment. Customer expectations emerged as a driver of IoT adoption as businesses need to make themselves appear as endeavouring to keep up with the pace of technology. Therefore, businesses might feel pressured to introduce IoT so as not to be left behind, as expressed by one respondent.

*There's a fantastic opportunity for the first UK high-street retailer who uses IoT and uses it well, a massive opportunity for such retailer to use it commercially as a USP [unique selling point]. You should be at the forefront of IoT and driving it. You shouldn't be just a follower; you should be a trendsetter, a champion spearheading it.* (Respondent 7)

The salient part of the above statement, however, is the call for IoT to be 'used well' in a retail environment to the extent that it can be considered the USP of a brand. While IoT is now increasingly being used in the service industry, a business firm must be able to make itself synonymous with IoT at least in the UK context.

The respondents also identified the potential to develop personalised customer journeys through a store from the vast amount of personal data gathered by the connected devices. However, they also took a higher-level approach, noting how IoT devices, which are generally smaller and cheaper, enable a business firm to effect fast change in a cost-effective way, as can be seen in the excerpt below.

*IoT is related to the concept of a device that fits very neatly with a lot of the paradigms that businesses tend to go for nowadays, such as 'run fast and fail quickly' for better or for worse….* (Respondent 8)

The challenges noted were also diverse, possibly due to the different specialisations of the respondents. However, most of them focused on the heterogeneous nature of IoT devices, as demonstrated by the excerpt below.

*Most of the IoT stuffs have different specs. There are no standards to follow, so each of the devices is built on a different base, and having a baseline for the infrastructure is difficult because it cannot be defined.* (Respondent 1)

Further probing revealed the criticality of balancing security, costs and benefits and tackling some of the key challenges of IoT implementation, as the following excerpt shows.

*You have to be pragmatic and weigh what it [IoT] will deliver for your business against what security controls you need…. The controls you put on an IoT device need to be proportional to the stuff you're trying to protect.* (Respondent 9)

Along the aforementioned lines, several respondents (Respondents 3, 6, 7 and 9) indicated that because security is an operational cost rather than a revenue stream, businesses need to take care not to overspend for security (i.e. 'not spending £5 million on a £1 million problem' [Respondent 9]). This highlights the fact that for IoT-enabled stores and spaces, security is an ongoing cost that will need to be budgeted for continuously; as such, the cost for it has to be proportional to the value of what is being protected (e.g. central corporate systems vs point of sale) to avoid overspending.

### 4.2 Devices and Infrastructure

Security is considered both at the level of the individual device and at the level of the network. With regard to the security features of a device, provenance is of particular concern, especially in relation to the different regulations concerning data handling. For example, one respondent (Respondent 9) observed how the geopolitical environment could create obstacles for security. He noted that since the US ban on Huawei products, for example, Huawei tablets have no longer been on Microsoft's list of enterprise devices. The implication of this is that companies can no longer patch and maintain Microsoft applications on Huawei devices and are thus potentially vulnerable to attacks; in other words, companies can no longer use these tablets in a secure way for IoT service environments.

Given the low cost and ease of use of many IoT devices, many employees feel empowered to introduce such devices into the store network. A key challenge here is keeping track of the many different devices on the network, and especially of 'shadow IT [information technology]' (Respondent 2), IT devices or systems used within an organisation without explicit IT department approval. Respondent 7 expounded on this by saying that the organisational structure exacerbates the risk posed by the use of 'shadow IT' because for projects that are not considered big enough to involve the IT department, the security function tends to be bypassed. The accessibility of IoT devices increases the opportunities for using shadow IT and amplifies its negative impacts.

With regard to practical advice regarding the implementation of IoT within a store, the advice that was given by the respondents was to ensure consistency (e.g. 'try and have some consistency' [Respondent 8]). Given the security challenges introduced by a heterogeneous environment, this idea is based on the premise of 'simplifying the problem so that remote control would be easy' (Respondent 5). Simplicity and homogeneity suggest some advantages, as shown below.

- Tracking of devices
- Monitoring of threats and alerts for particular systems
- Easier training
- Identification of experts for all devices
- Faster response in a crisis, with all the devices following the same protocol
- Reduced operating costs

While securing one thing well does potentially make vulnerability common, on balance, it seems more viable than successfully protecting and monitoring multiple systems.

## 4.3 Project Management

Management of IoT implementation within a customer-facing service environment emerged as one of the important themes from the study data. Table 3 provides an amalgamation of the suggested steps, from concept to rollout.

Table 3. Suggested steps for managing the implementation of Internet of Things (IoT) systems

| Step | Respondent | Detail |
|---|---|---|
| Business case | 1, 3, 4, 7, 8, 9 | Why choose an IoT solution? What data will be collected? What business advantage will be gained? What is the effect on the staff/customers? What is the budget? How much will it cost? |
| Project design | 3, 5, 7, 8 | Do you need to hire anyone to ensure you have the right knowledge? What is the project environment? Can you make changes to the building? Does the solution meet the business requirements? Does the solution meet the security requirements? Does the solution satisfy internal compliance? |
| Procurement | All | Soft market research on the best provider. What is the built-in security of the devices? Does the vendor sufficiently protect data? Is the vendor lawful and ethical? |
| In-house device test | 3, 6 | Take the device apart to see exactly how it works. Does it function as promised? |
| Closed trial | 7, 8 | If possible, test in a mock environment to see if it will interfere with the existing processes without causing revenue loss. |
| Real trial | All | Trial in a variety of store types. Are there any security flaws? Are there any usability flaws? Iteratively optimise the system. Does it meet the business case? |
| Service design | 3, 6, 7, 9 | Who is responsible for the maintenance of the devices, for both functionality and security? What does that involve? What are the running costs? |
| Rollout | All | Batch approach. It is suggested that rollout be started near locations that can be easily accessed by the maintenance personnel. |
| Lifecycle monitoring | 3, 5, 7, 8, 9 | Functional delivery of 'service design'. Periodically assess against the business case – Is it meeting the objectives? |

Regarding security, it is important to think about it from the start of the project as retrofitting a security system with the other systems typically incurs higher costs than building it in before the start of the project (Respondent 5). A respondent pointed out another need, as shown below.

*There should be at least a partial InfoSec input at each stage. Some can just be a chat facility and some can be far more in depth. But if you have one at every stage, you'll have end-to-end assurance.* (Respondent 8)

However, security is a dynamic concept, as expressed by a respondent in the following interview transcript excerpt.

*The security posture of a device is never static. Things change; people are always looking for ways around things. It might be secure on day 1, but it might not be secure down the line.* (Respondent 3)

The aforementioned concern perhaps presents more of a challenge for IoT devices than for other technologies because many IoT devices do not have a direct user interface. It is not always obvious (e.g. through alerts) when a device needs to be updated or when its license is about to expire. Thus, business and security requirements will need to be constantly monitored and assessed (project design). Such potential changes will need to be reflected in the procurement costs and service design (running costs).

Another approach to addressing project management issues was raised by Respondent 9, who commented that 'issues tend not to be with technology; they are more regulatory and thus related to HR [human resources], to people's ability to adapt to change'.

Although common to all forms of organisational change, many IoT implementations exhibit an observational nature. Organisations are likely to face complaints from their employees regarding privacy infringement, whereby the workplace becomes a 'Big Brother state' (Respondent 9). The implication of this for IoT implementation is that it is likely to be received with scepticism and resistance by the employees, which in the longer term may have negative consequences for the project in general.

### *4.4 Privacy and Data*

When discussing the challenges of implementing IoT solutions, the respondents expressed that they consider data and privacy almost synonymous with IoT. One could argue that data and privacy are more universal concerns, extending beyond the area of IoT (e.g. in online social networking applications such as Facebook). Many of the respondents expressed certainty that their organisation would be General Data Protection Regulation (GDPR) compliant. However, there was also a concern that organisations do not always fully understand or know the volume and contents of the data they are capturing.

With IoT's capacity to monitor an environment with higher granularity than before, it is not so much the individual data streams but the collective context of the data that pose an issue. As Respondent 5 said, 'Supposedly anonymised metadata can be so detailed that they can be easily used to identify the source.'

By not understanding the power of combined data, organisations fail to build sufficient protection mechanisms around them. However, cybercrimes are not new crimes and are not always facilitated by state-of-the-art technology. It is the crossover between the two that is new. This is what makes the versatility and omnipresence of IoT more of a challenge: it introduces an exponential increase of threat vectors that can be combined, as expressed by one respondent.

*IoT is reasonably new…. It always takes people's skills and experiences to develop controls to catch up with a technology…. When you mix a niche technology area with a fast-growing new concept, they catalyse each other in terms of risk. You don't know what to look out for, and you don't know how to secure it even if you did!* (Respondent 8)

With regard to the implications of the foregoing for stores, privacy is subject to the practical bandwidth limitations of businesses and their processes. For example, Respondent 10 commented, 'A company should try to make sure that only relevant data are taken from systems. We should cherry-pick what is useful and send it to the cloud.'

The foregoing is particularly salient when, as pointed out by Respondent 7, one considers the store environment, where the bandwidth is limited and has to be shared with point-of-sale devices that are critical for business operations. It is also relevant for the scenario where a personalisation service is available in the store but a customer decides to disable such function and the request to stop data collection and notifications needs to be processed quickly to respect such customer's right to privacy.

Another issue is related to the concept of data security in the supply chain. Within a service environment, there may be a disparity among service providers in prioritising data protection, with some providers being 'very careful with their customers' personal identifiable information as it's their lifeblood' (Respondent 5) and with hardware device manufacturers prioritising data protection less. Respondent 5 added the comment below.

*You need to know that a third party has the same or better security controls than you have over data because while you're outsourcing the processing of your data, you're not outsourcing the responsibility for them.*

Due to the foregoing, it is critical to have the right of audit over the supply chain.

### 4.5 Keeping Humans in the Loop

The most typically discussed topic was that of usability versus security. Respondent 8 said, 'There's the old InfoSec joke that the most secure computer is the one that's turned off, but that's no use to anyone!'

Respondent 9 shared anecdotal evidence from a large retailer who had changed its policy from not allowing the use of any personal device on the shop floor to allowing all employees carrying a mobile device to look up products and to inform customers where these can be located. The strategy is based on the premise described below.

*Security has to be very balanced. If you make something so controlled that people can't use it, they won't! They will do something completely different. And considering how the world is today, there will always be a different way of doing something.* (Respondent 9)

In effect, while retailers may not have much control over security, they can be confident that all their employees are at least using the same process, thereby limiting the unknown variables, which then makes it easier to place security controls over the process.

When talking about the 'humans in the loop', Respondent 3 commented, 'It's an interesting debate, and there are many things to consider with regard to whether you should try to protect against humans or try to educate them. As for us, we try to do both.' In relation to this, all the respondents said that education on security should revolve around generic good practices, but some respondents gave more practical suggestions, such as making the message clear and succinct, relating the education to the employees' responsibilities and using multiple sources to engage the employees.

## 5 Discussion

### 5.1 Synthesis of the Results with the Existing Literature

This study aimed to obtain a better understanding of the drivers and challenges of IoT implementation in the retail industry, particularly in smart stores, utilising the SD logic and the DOI approach. Our data analysis showed that when implementing IoT in a store, five critical aspects have to be taken into account: business considerations, devices and infrastructure, project management, privacy and data and keeping humans in the loop. This section integrates our findings into the existing relevant literature and further discusses the security-related implications for each of these five themes.

*5.1.1 Balancing drivers and challenges*

IoT's power to generate unprecedented insights into customer behaviour is widely reported as a driver of IoT adoption in the literature (Gregory, 2014; Lee & Lee, 2015; Metallo et al., 2018). The ability to increase the efficiency of processes through the automation of menial tasks is also commonly acknowledged. For the service sector, in particular, there is a trend towards personalisation and experience-driven sales (Accenture, 2017; Balaji & Roy, 2017; Gregory, 2014; Grewal et al., 2017) because these are expected by customers (Priporas et al., 2017).

However, we found in our study that responding to competitive forces is likely a stronger argument in favour of implementing IoT in physical stores than satisfying customers is. That is, while retailers suggest that they do experience pressure to leverage IoT to meet their customers' expectations, more often than not they perceive IoT-enabled solutions as critical for achieving a competitive advantage and maintaining it. At the same time, our study showed that IoT-based systems help businesses improve the customer journey because they support personalisation and can obtain customers' insights, the latter often being better and more affordable than competing market research solutions.

Nevertheless, investing in technology is not as straightforward as choosing to invest in IoT because one's competitors are doing the same, which can result in more challenges than opportunities. Device heterogeneity is widely considered the key obstacle to businesses' IoT adoption (Lee & Lee, 2015; Sicari et al., 2015) as this entails additional complexity, with implications for security and interoperability. Specifically, while heterogeneous IoT devices monitor the environment and produce data streams in multiple formats (e.g. video, sound and metrics), they obtain a vast quantity of diverse data, with each format requiring a different transmission and storage infrastructure to be considered 'secure' (Roman et al., 2011). In terms of the use of the obtained data for data analytics, having such a large pool of data poses a great challenge for data storage and relevance (Sun et al., 2016). This relates to the adage 'bigger data are not necessarily better data' (Lazer et al., 2014), which supports the process of pre-filtering 'useful' data. Our findings thus lend further importance to the bandwidth ceiling in a store environment and to knowing what data are being generated to be able to manage them effectively.

In addition, device heterogeneity relates to the security of IoT systems by default. We found out that in the retail sector, the security system has to be proportional to what it is protecting. That is, decision makers need to identify the minimum security expenditure that can satisfy retailers' risk appetites. Proportionality will allow businesses to make reasonable expenditures, which will in turn allow such businesses to maintain normal operations. The implications of the existence of multiple heterogenous devices for security costs have been discussed in earlier studies (Gierej, 2017; Zhao & Ge, 2013), but without referring to proportionality. This may be because IoT studies are typically not multidisciplinary and adopt either a business approach (e.g. Metallo et al., 2018) or a security-focused approach (e.g. Ning et al., 2013). As such, the business side sees security as something that needs to be addressed but not necessarily rigorously while security research is focused on best practices, unencumbered by practicalities.

When considering the costs, Yee (2004) notes the importance of specificity in parameterising system requirements and capabilities. This, he explains, facilitates more accurate security cost projections because specificity supports demarcating the necessary spending on different types of systems to make them secure on the basis of their criticality for the business (in terms of function or data). Our findings show that even if this observation was made before IoT became popular, it remains pertinent for ensuring the financial viability of security.

*5.1.2   Security of heterogeneous devices and homogeneity*

With regard to devices and infrastructure, our findings show that having little to no control over the built-in security of devices manufactured by a third party can be problematic. On the one hand, this is related to the issues arising from the heterogeneity and complexity of devices and infrastructure. On the other hand, it also refers to standardisation and the lack thereof. The lack of standardisation suggests increased operational costs because it requires the involvement of several different experts in training or recruitment, who will need to manage and monitor the numerous but heterogeneous devices and systems.

Various governments have identified that third-party device manufacturing and the lack of standardisation of such devices' security features are indeed critical. For example, in 2019, the UK government released a list of 13 'secure by design' features that all IoT devices should have to ensure a minimum level of security (ETSI, 2019). In addition, the US National Institute of Standards and Technology released two reports addressing IoT device manufacturers (Fagan et al., 2019a) and the businesses and organisations that use IoT (Boeckl et al., 2019).

Our findings suggest that homogenising the device environment when possible can reduce the security challenge and provide intuitive customer-facing services. Nevertheless, homogenisation of the device environment does not appear prominently in the security literature to date. Boeckl et al. (2019) suggest that devices from the same manufacturer are easier to manage and monitor centrally whereas devices from different manufacturers introduce vulnerabilities in the IoT lifecycle and require diversified management of updates and alerts. The latter leads to further complications, whereby integrating a wide range of devices within a single information security policy may result in overload and uncertainty, leading to inconsistent adherence to policies. For this reason, standardisation is said to reduce the need for a diversified security policy (D'Arcy et al., 2014). However, our findings show that whether to 'put all your eggs in one basket' and seek to improve manageability results through an acceptable trade-off is ultimately a business decision. On this basis, appreciating which baseline security features should be included in an IoT system is critical for procurement decisions.

*5.1.3   Managing Internet of Things implementation*

Our study showed that the management of an IoT project influences its implementation, the security lifecycle and the organisational culture. Implementation of the IoT infrastructure in a store should always start with a business case, such as the typical information system projects, and should consider the customer journey. This approach allows the development of a common understanding of the IoT system and its return on investment (ROI). Specifically, Palattella et al. (2016) explain that with regard to IoT, there are three main areas with high ROI: efficiency savings, big data (often considered the main investment motivation due to the superior customer insights they provide) and infrastructure costs (because the use of IoT reduces the cost of rewiring the store space).

In addition, our findings show that the differences between IoT and the 'traditional' IT (modes of interaction with the physical world, access, storage, monitoring, security and privacy and functionality) (Boeckl et al., 2019) require close consideration during the entire lifecycle of an IoT project (technical and service design, procurement, implementation and monitoring). Along these lines, Fagan et al. (2019) offer some recommendations, as shown below, which we consider relevant to IoT projects in the retail sector particularly in relation to procurement.

1) What are the security features of the device?
2) What exactly does the device do and what mechanisms does it use to facilitate that?
3) How are software and firmware updates delivered?
4) When does the device stop receiving product support?
5) How should the device be handled at the end of its life?

Our study showed, however, that security considerations need to go beyond procurement costs or should not be confined within the project lifecycle. Instead, security should be approached as an ongoing running cost and one that is considered together and iteratively with usability and the customer journey so as to reduce its potential conflict with these two and to prevent spiralling costs. This notwithstanding, the idea of making a system both more secure and more efficient is often considered a 'unicorn state' in InfoSec (Respondent 8).

With regard to organisational culture, our findings show that the introduction of IoT systems often results in changes in everyday workflows and employees' roles. Earlier studies have explored the role of employees in relation to IT-induced transformation in the service sector while drawing attention to how technology may either 'augment' employees' capabilities or replace employees altogether to remove inherent human performance variability (Larivière et al., 2017; Pavlou, 2018). The latter, however, was not confirmed by our study. On the one hand, our findings suggest that there is a need for employees to be ready for change or to successfully engage with a changing role, which necessitates training opportunities, awareness-raising campaigns and change management programmes (Larivière et al., 2017). On the other hand, our findings show that the greatest concern with regard to IoT adoption seems to be employees' perceptions of the effect of IoT adoption on their privacy, which can negatively influence the organisational culture. In addition, one could argue that uncertainty and potential negative perceptions of one's privacy in the workplace may create risks for the organisation's security. Our study does not offer direct evidence for this, but the existing literature shows that disgruntled employees often pose an insider threat to an organisation's security (Greitzer et al., 2012).

*5.1.4 Privacy concerns and the customer journey experience*

To date, concerns regarding data privacy and trust have been hindering the widespread adoption and diffusion of IoT (Palattella et al., 2016). These concerns have also been expressed in our study, but from a different perspective. The participants in our study consider complying with GDPR the default position, which nevertheless creates considerable challenges with respect to handling data and specifically personal identifiable information (PII). They consider their organisations able to function responsibly throughout the process of collecting, storing and processing PII despite the involvement therein of third parties, whose practices cannot always be controlled. This suggests that they have a deep knowledge and understanding of the nature of data along the entire supply chain.

Interestingly enough, the participants in our study further highlighted that their customers would happily provide their personal information if they were to receive something in return, such as personalised services. However, customers may likely have little understanding of what such personalisation requires and what providing their PII to an organisation may entail (Walker, 2016). For example, personalised services such as targeted advertising may be perceived as too intrusive and may thus not be well received (Inman & Nikolova, 2017a). Such negative perceptions may be exacerbated by practicalities such as the bandwidth limitations within a store environment. In hypothetical scenarios where customers opt out of personalised notifications as they move around the store, a low bandwidth may result in a delay in processing their request. In this case, the customers may interpret such delay as the store's ignoring of their request and may think that their privacy is thus in jeopardy. Naturally, this will have negative implications for the customer journey. In addition, such delay may be non-compliant with GDPR, and on the basis of our findings, may indicate the stores' low level of responsibility regarding data and request handling. As a result, deep knowledge and understanding of PII is crucial, and this also extends to metadata. Specifically, our findings reflect the risks posed by privacy breach when metadata can be pieced together despite the previously applied anonymisation techniques. A recent report by the National Institute of Standards and Technology addresses this issue by recommending continuous mapping of PII data through a system to mitigate deanonymisation via data aggregation (Boeckl et al., 2019).

### 5.1.5  *Usability and security of Internet of Things in a customer-facing environment*

Our study showed that keeping humans in the loop relates to achieving a balance between security and usability, cultivating useful behaviours and training. We found that security protocols often restrict store employees and reduce their opportunities to innovate while serving customers. The participants in our study indicated that training (particularly security-focused training) is important to combat this, but such training can be beneficial only when it is relevant to the employee experience and specific to the devices typically used by employees as part of their workflow.

While the aforementioned balance between security and usability has been addressed by earlier studies (e.g. Ben-Asher et al., 2009; Yee, 2004), little to no attention has been paid to the IoT context thus far. Our findings show that IoT introduces further opportunities to circumvent security protocols. We also found that security systems need to be designed with human behaviour in mind to address potential usability shortcomings. Previous studies have found that when usability is low, users are inclined to 'bend the rules' and enact workarounds (e.g. Zamani et al., 2019), behaviours that are also relevant to the IoT context. In many cases, introducing an IoT infrastructure may result in increased complexity of the technological environment and may introduce intricacies in employees' workflows. Studies have shown that in such cases, increased security requirements may result in security-related stress (SRS), which has been associated with moral disengagement and security policy violations (D'Arcy et al., 2014). It could also be argued that IoT could streamline processes and could therefore reduce rather than increase complexity, with frontline employees needing to abide by fewer security protocols and thus having reduced SRS. In all cases, however, employees will more often than not be able to find ways to work around the system (Alter, 2014) and to breach security protocols.

### 5.2  *Implications for Theory*

Our field study provided a content-rich understanding of the specific drivers and challenges of IoT adoption and diffusion in smart stores. It also provided an enhanced understanding of the initial relevant theoretical aspects, further specifying IoT technology in-store. Our discussions with the study participants were focused on our initial research question: What are the drivers and challenges of IoT implementation in smart stores, and which of them are relevant to customer-facing services? Therefore, the major contributions of our study pertain to the security and privacy domains, which are major concerns for customers and retail stakeholders and challenges for successful IoT diffusion. Specifically, we found that privacy and security could jeopardise the customer journey and could have negative implications for customers' buying behaviour and challenges for smart stores implementing IoT.

The second contribution of this study is that it extended the prior research on customer experience and satisfaction from an IT perspective. Specifically, while the prior work focused on smart customer experience in the retail domain primarily from a consumer perspective (e.g. Roey et al., 2017), our study extended the current understanding of IoT in particular while considering the challenges it poses for both customers and retailers. This is of particular significance because the IoT technology offers specific opportunities but also poses specific challenges. On the one hand, the opportunities offered by IoT and the challenges posed by it are distinct from those offered and posed by other technologies, such as blockchain. On the other hand, to date and to the authors' knowledge, the literature pertaining to customer satisfaction has not yet identified how these opportunities and challenges may influence consumers but has also not yet identified their implications for retailers seeking to offer unique customer journeys.

In addition, because our study focused on IoT implementation in stores as our unit of analysis, with a view to exploring customer perceptions of it, we adopted the principles espoused by retail scholars to identify the specifics of the customer experience (Homburg et al., 2017). That is, in conceptualising our study, we adopted the innovation perspective, which is often applied in studies on smart store technologies (Pantano & Viassone, 2014). We adopted the conceptual lens of the DOI theory (Rogers, 2003), which determines the rate of adoption of innovations. To date, technology adoption theories such as the unified theory of acceptance and use of technology (Venkatesh et al., 2003) and the DOI theory are often used for theoretically framing studies in the area of smart retail technology, with a view to comparing customer behaviour and switching behaviour between technology products (e.g. Kamolsook et al., 2019). Leveraging the DOI theory offered us a theoretical lens for exploring the IoT technology in customer-facing environments. However, we integrated this with the SD logic (Vargo & Lusch, 2004) to define and account for the service aspects. We consider this an important contribution. Our study offered evidence that combining the DOI theory with an SD logic perspective can result in a more holistic understanding of the adoption and use of IoT by retailers considering both the technical and service features of the solution. This is an important implication as there are numerous calls for research on the use of IoT in the retail sector and for the purpose of exploring smart technologies and their impacts on service and service innovations (Roy et al. 2017).

### 5.3 Implications for Practice

The findings of the present study have important implications both for theory and practice by providing new knowledge on the concept of implementing IoT in a customer-facing environment using a pragmatic approach. To the best of our knowledge, this was one of the first studies that approached IoT security within the context of IoT implementation in a retail environment. Through our findings, we enhanced the existing IoT literature (Whitmore et al., 2015; Li et al., 2015), offering a richer understanding of security in smart stores informed by the SD logic, IoT network security and the DOI theory.

The study further offered a comprehensive understanding of the practical considerations that companies need to make when implementing security for IoT, such as the trade-off between security and costs. We offer a range of practical solutions and recommendations for the implementation of a customer-facing IoT system in-store. Among these, we consider the business case, which is often a factor for traditional IT projects, the most critical (Kappelman et al., 2006). Companies should implement IoT either to solve a business problem or to access a specific benefit rather than simply because they 'want more technology'. This information should provide the basis for a business case that identifies the expectations of the project. While the usual InfoSec rules apply, for IoT in particular, companies should map out the data flows to identify the streams that require higher levels of protection and to quantify them against a company-wide standardised scale.

With regard to data and devices, these should not be considered in isolation because the aggregate power of IoT data makes it as dangerous as it is useful. Equally, data collection for facilitating personalisation and analysis is very important for businesses. Indeed, the power of IoT lies in its ability to contextualise the shop environment at an unprecedented level of granularity. Streamlining the collected data reduces the demand for storage and transmission, both of which require finite resources. In doing this, however, retailers need to ensure that sensitive data are suitably encrypted when stored or transmitted. This poses a challenge as the demarcation between 'sensitive' and 'non-sensitive' data gets blurred when data are collectively aggregated. This is an increased concern for businesses considering setting up IoT-enabled store environments, which will need to ensure that their security protocols and controls extend across their supply chain, including their device suppliers and manufacturers. Further, the retail sector is characterised by complex supply chains, whereby device suppliers and manufacturers often collaborate with hardware and software suppliers located in different countries, where the data protection regulations may differ considerably. Therefore, businesses will need to consider the potentially severe implications of non-compliance with the local data protection regulations due to country-level inconsistencies.

### 5.4  Limitations and Future Research Direction

As with all empirical studies, the current study had inherent limitations. Our study was an exploratory field study, and we collected empirical data through semi-structured interviews. While our findings shed light on the implementation of IoT in the service industry, they cannot be generalised without caution. Primarily, we propose their validation within a similar context, and from there, their extension to theory, as is often the case with qualitative studies, which can in turn be validated and extended to different contexts (Davison & Martinsons, 2016, p. 247). In addition, we would welcome further research into the concept and effects of homogenising the device infrastructure both in terms of risk and quantifying benefits, which we think is essential but is outside the scope of our study. Although the concept of homogenising devices is briefly examined in this paper, the validity of the concept must be established through further research to determine if additional risks will be introduced by less variance in IoT devices. Furthermore, in the present study, security experts from different retail companies were interviewed regarding the drivers of and security challenges posed by IoT implementation in the retail industry. Future research should aim to capture the opinions and perspectives of marketing and customer insights professionals working in the retail industry using IoT to better understand customer security concerns.

Another interesting avenue for further research is exploring what makes a store environment synonymous with the concept of IoT; how IoT may be implemented together with other advance technologies, such as blockchain, both for payments and for security purposes and how these combined technologies would influence customer and employee expectations in relation to security design and privacy expectations. In a world where businesses are expected to use information and communication technologies ahead of their competitors and be trendsetters, it would be interesting to see whether and to what extent an equal amount of care is expected or applied in making this offering secure by design. Previous studies have argued, for example, that in some cases blockchain-powered systems can provide an additional layer of security (Zamani et al., 2020). A similar study could further focus on which types of IoT devices are most used in the retail sector, the particular security problems these devices bring about and whether complementary technologies can address the perceived and real security issues. Finally, we note that in our study, we explored the challenges and especially the security and privacy concerns not only of IoT-implementing organisations but also of customers. As such, we did not examine aspects of performance. However, performance is highly linked with the identified challenges as any of these can directly affect performance. Future studies could explore this link through a survey to identify, measure and explain the effect of security on IoT implementation and its direct/indirect links to performance.

# 6 Conclusion

This study aimed to explore the drivers of, and security challenges posed by IoT implementation in a store, focusing on the retail service industry sector. The retail industry is moving towards the provision of personal experiences to their customers. IoT can facilitate this by sensing and gathering data from a store environment and streamlining the existing processes. Furthermore, it can be used as a cost-effective and expeditious way to effect the required change. At the same time, the use of IoT can present a myriad of opportunities for offering new services, many of which customers already expect and there is already a market pressure for. However, there are many challenges regarding enhancing the customer experience and addressing the relevant security and privacy concerns on the part of businesses. Future studies need to carefully consider these challenges before, during and after the implementation of the IoT system.